\documentclass[10pt,twocolumn]{article}

\usepackage{graphicx}
\usepackage[utf8]{inputenc}
\usepackage{amsmath}

\usepackage{algorithmic}
\usepackage{algorithm}
\usepackage{subcaption}

\newtheorem{definition}{Definition}

\begin{document}
%

\title{Processing Regular Path Queries on Arbitrarily Distributed Data}
%
%
%
%
%

%
\author{
Alan Davoust and Babak Esfandiari\\
       Dept of Systems and Computer Engineering\\
       Carleton University\\
       Ottawa, Ontario, Canada\\
       \{adavoust,babak\}@sce.carleton.ca
}
\date{}
\maketitle

\begin{abstract}
Regular Path Queries (RPQs) are a type of graph query where answers are pairs of nodes connected by a sequence of edges matching a regular expression.

We study the techniques to process such queries on a distributed graph of data.

While many techniques assume the location of each data element (node or edge) is known, when the components of the distributed system are autonomous, the data will be arbitrarily distributed.

As the different query processing strategies are equivalently costly in the worst case, we isolate query-dependent cost factors and present a method to choose between strategies, using new query cost estimation techniques.

We evaluate our techniques using meaningful queries on biomedical data.
\end{abstract}




\section{Introduction}

Regular Path Queries (RPQs) were first introduced as part of a query language for graph databases~\cite{cruz1987graphical,Consens:1990:GVQ:93597.98748}, and gained particular interest as a way of querying the distributed graph formed by the World Wide Web pages and hyperlinks~\cite{abiteboulvianuqueriesandComputationWeb,abiteboul1997regular,DBLP:journals/jodl/MendelzonMM97}. More recently, there has been renewed interest in RPQs with the development of the Semantic Web: a form of RPQ was included in version 1.1 of the SPARQL query language (``property paths").

In this paper, we study the problem of processing regular path queries over distributed data.

Several distributed query processing algorithms were proposed for the graph of Web pages and hyperlinks~\cite{abiteboul1997regular,DBLP:conf/icde/FernandezS98} and for distributed graph databases~\cite{DBLP:journals/tcs/ShoaranT09}. These algorithms rely on a key property of these two settings, which is that the nodes in the graph are \emph{localized}, in the sense that identifiers of nodes can be directly mapped to a location in the network where the node (i.e., a description of the node and of its incident edges) will be found.

However, when the components of the distributed systems are \emph{autonomous} and can freely choose which data they host (e.g. peer-to-peer (P2P) database systems or Semantic Web servers), this property does not hold: instead, data resources may be found in arbitrary locations, and may also be replicated in multiple locations. 

In fact, even when data can be described as localized, a difficulty occurs if we want to use the ``inverse" operator\cite{DBLP:journals/sigmod/CalvaneseGLV03,DBLP:conf/pods/Baeza13}, which is used to express path expressions where edges are followed in the ``reverse" direction. In the Web context for example, traversing a hyperlink in the reverse direction would require finding all the pages that link to a given page, which requires searching the entire Web.

In other words, for \emph{Regular Path Queries with Inverse} on distributed data, data can be considered \emph{localized} only if the distribution and placement of data is controlled in a centralized way, for example by a shared algorithm. 

\begin{figure*}[htb]
 \centering
 \begin{subfigure}[b]{0.33\textwidth}
  \centering
 \includegraphics[scale=0.5]{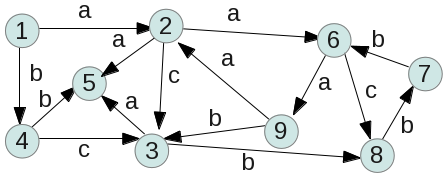}
\caption{An example graph of data.}
\label{fig:exgd}
 \end{subfigure}
 \begin{subfigure}[b]{0.63\textwidth}
\centering
\includegraphics[scale=0.46]{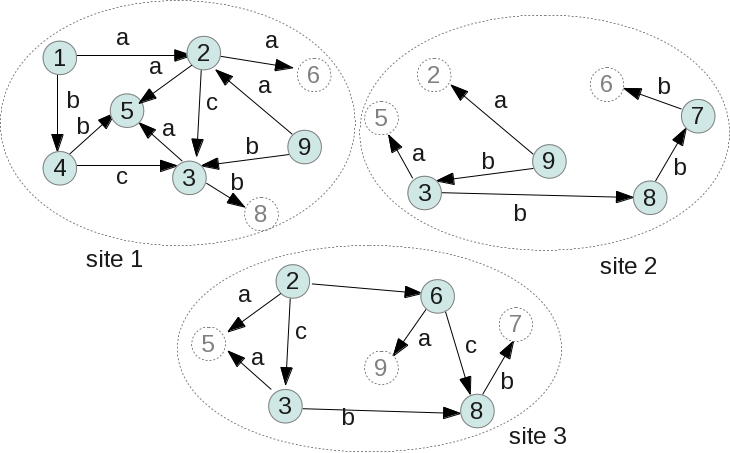}
\caption{A graph of data arbitrarily distributed and replicated between 3 sites.}
\label{fig:exgd-DS2}
 \end{subfigure}
\end{figure*}

In every other case, the data should be considered as \emph{non-localized}. This setting of \emph{non-localized} data is illustrated in figure~\ref{fig:exgd-DS2}. 
 
%
%

We show that existing distributed query processing techniques proposed for localized data are unapplicable to this setting, and evaluate centralized techniques, based on the idea of dynamically accessing the sources for each query, either in a single step before executing the query locally, or else on the fly during query execution~\cite{hartigsparql2009,Harth:2010:DSO:1772690.1772733}. Ladwig and Tran~\cite{Ladwig:2010:LDQ:1940281.1940311} refer to the former approach as ``top-down", and to the latter as ``bottom-up".

%

In the worst case, both strategies may result in retrieving the entire graph of data, which is unmanageable in practice. However, in practice we can expect the majority of queries to be more \emph{selective}, and ideally we would like a query processing strategy that performs well on ``real-life" cases, and we would also like to determine conditions to choose between the different strategies, and identify \emph{a priori} the problematic cases. 

We show that the optimal strategy choice depends on the selectivity of the query, and propose two techniques to estimate the ``selectivity" of queries based on a sample of the data. We evaluate our techniques against a real dataset from the biomedical domain and a set of meaningful queries.

The rest of this paper is organized as follows. In section \ref{sec:background} we provide some background definitions and algorithms for RPQ in general. In section \ref{sec:distrib} we evaluate existing query processing techniques for the specific case of non-localized (arbitrarily distributed) data. In section \ref{sec:discriminant}, we evaluate and compare two query execution strategies against the biomedical data, and determine conditions to choose between the strategies. We present our query cost evaluation techniques in section \ref{sec:costeval}. Finally, we illustrate the use of all these techniques in a brief case study (section \ref{sec:scenario}), and draw some conclusions in section \ref{sec:conclusion}.

\section{Definitions and Algorithms for RPQs}
\label{sec:background}

Informally, the idea of a regular path query is to find pairs of nodes in a graph of data, such that the path (sequence of edges) from one node to the other matches a given regular expression. Two variations of RPQs have been considered in the literature: multi-source queries, that return every pair of nodes matching the query, and single-source queries, where a single ``start node" is given. 

We now define these queries more formally.

\subsection{Notations}

All queries are applied to an edge-labeled directed graph $G_D = \langle V, E \rangle$, where $V = \{v_0, v_1,\dots, v_N\}$ are nodes and $E \subset V \times \Delta \times V$ are edges labeled from a set of labels $\Delta = \{\delta_i\}$. 

A path in $G_D$ from a node $v_0$ to a node $v_k$ is a sequence of adjacent edges $(v_0, \delta_1, v_1),(v_1, \delta_2, v_2),\dots,$ $(v_{k-1}, \delta_k, v_k)$ starting at node $v_0$ and ending at node $v_k$.

The notation $v_0 \overset{w}{\rightarrow} v_k$ indicates that there exists a path from $v_0$ to $v_k$ such that the sequence of edge labels $\delta_1,\delta_2,\dots,\delta_k$ along this path forms a word $w$. If $r$ is a regular expression, we note $L(r)$ the regular language defined by $r$.

\subsection{Definitions}

\begin{definition}[Multi-source Query]
A multi-source query $Q_r$ is defined by a regular expression $r$ over $\Delta$. When $Q_r$ is applied to the graph $G_D$, the answers to $Q_r$ are defined as follows:
\begin{equation*}
Ans(Q_r, G_D) = \{ (v_i, v_j) \in V \times V | v_i \overset{w}{\rightarrow} v_j, w \in L(r)\}
\end{equation*}

\end{definition}

\begin{definition}[Single-source Query]
A single-source query $Q_{r,v_0}$, applicable to the graph $G_D$, is defined by a regular expression $r$ over $\Delta$, and a distinguished node $v_0$ of $G_D$. The answers to $Q_r$ are defined as follows:
\begin{equation*}
Ans(Q_{r,v_0}, G_D) = \{ v_j \in V| v_0 \overset{w}{\rightarrow} v_j, w \in L(r)\}
\end{equation*}

\end{definition}

\subsection{Regular Path Query With Inverse}
\label{sec:rpqidef}

The notion of a Regular Path Query can be extended to the notion of a Regular Path Query with Inverse (RPQI~\cite{DBLP:journals/sigmod/CalvaneseGLV03}, also sometimes called 2RPQ~\cite{DBLP:conf/pods/Baeza13}), when directed edges can be traversed in both directions. Such queries can be particularly useful in directed acyclic graphs, since for any pair of nodes related by a path $v_1 \overset{w}{\rightarrow} v_2$, there is no path from $v_2$ back to $v_1$. Using the ``inverse" operator allows an extended notion of a path to exist between $v_2$ and $v_1$.

Conceptually, adding the inverse operator amounts to duplicating the edges in the graph, adding, for each existing edge, a parallel edge pointed in the opposite direction. The additional edges semantically represent the inverse relation of those represented in the original graph. We construct the labels of the new edges in a systematic way, by appending a ``$-1$" superscript to the existing labels.

Formally, we consider the extended alphabet $\Delta'$ defined by extending $\Delta$ with an additional symbol $\delta^{-1}$ for each symbol $\delta$ in $\Delta$. 

For a given graph of data $G_D$, we construct the extended graph $G'_D =\langle V, E' \rangle$, where $E'$ is defined as: \begin{equation*}E'= E \cup \{(v_j,\delta^{-1}, v_i) | (v_i,\delta, v_j) \in E\}\end{equation*}

RPQI applied to $G_D$ are then simply RPQs on $G'_D$.

\begin{definition}[Multi-source RPQI]
A multi-source RPQI query $QI_r$, applicable to the graph $G_D$, is defined by a regular expression $r$ over $\Delta'$. When the query is applied to the graph $G_D$, the answers to $QI_r$ are the answers to the RPQ $Q_r$ applied to $G'_D$.
\begin{equation*}
Ans(QI_{r}, G_D) = Ans(Q_{r}, G'_D)
\end{equation*}

\end{definition}

Single-source RPQI are defined in a similar way.

\subsection{Example}

\begin{figure}[htb]
\centering
 \includegraphics[scale=0.5]{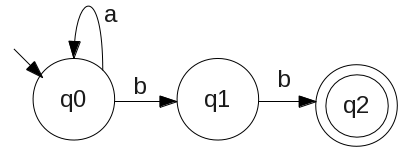}
\caption{A query automaton.}
\label{fig:exgd-automaton}
\end{figure}

As an example, we consider the graph of data shown in figure~\ref{fig:exgd}, and the following example queries, applied to this graph of data:

\begin{itemize}
 \item $Q1 =$ (1, $a^*bb$) is a single-source query that requests all nodes that can be reached from node 1 by a path matching the regular expression $a^*bb$. An automaton for this regular expression is shown in figure~\ref{fig:exgd-automaton}. The answers to the query are nodes 5 (path 1-4-5, $bb$) and 8 (path 1-2-6-9-3-8, $aaabb$). For node 8, there exist an infinite number of paths yielding this result, including any number of traversals of the cycle 2-6-9-2.
 \item $Q2 =$ ($ac(a|b)$) is a multi-source query that requests all pairs of nodes related by a path matching the regular expression $ac(a|b)$. The answers to this query are the pairs of nodes (1, 5), (9, 5) (both paths labeled $aca$), (1, 8), (9, 8),  (2, 7)(paths labeled $acb$).  
 \item $QI3 = $(1, $a^*b^{-1}$), is a single-source RPQI, where $b^{-1}$ denotes following an edge labeled $b$ in the opposite direction. Answers to $QI3$ are nodes 4 and 7 (paths 1-2-5-4 and 1-2-6-7). Again, the cycle 2-6-9-2 can be added to both of these paths.
\end{itemize}
 
\subsection{Basic RPQ Algorithm}
\label{sec:PAA}

The main algorithm to answer such queries (detailed below) was sketched in a 1989 paper by Mendelzon and Wood~\cite{DBLP:conf/vldb/MendelzonW89}.

We detail their general algorithm, which is the basis of most approaches (as we will show later on). We will refer to this algorithm as the \emph{product automaton algorithm}, or PAA:
\begin{enumerate}
\item build a finite automaton $A_1$ associated with the regular expression $r$. The initial state of $A_1$ is $q_0$, the accepting states are $\{qf_i\}$.
\item consider the graph of data as an automaton $A_2$ (nodes $\rightarrow$ states, edges $\rightarrow$ transitions), and compute the cross-product of the automata $A_p=A_1 \times A_2$.
\item 
\begin{itemize} 
  \item \textbf{(single-source RPQ)}: the initial node in the graph is set $(N_0)$: search $A_p$ from the initial state $(q_0,N_0)$ to find all reachable accepting states $(qf_k,N_j)$. All nodes $(N_j)$ are answers to the single-source query.
  \item \textbf{(multi-source RPQ)}: we are looking for all pairs of nodes related by the regular path: search $A_p$ from all initial states $(q_0,N_i)$ to find all reachable accepting states $(qf_k,N_j)$. All pairs of nodes $(N_i,N_j)$ are answers to the multi-source query.
\end{itemize}
\end{enumerate}

For step 3, any graph search algorithm can be used, such as breadth-first or depth-first. 

\subsection{Application to RPQI}

Since RPQI query answering can be reduced to answering RPQ on another graph, the same algorithm can be used. However, the graph $G'_D$ is an artifact of the definition and method, and is not materialized as a database. Traversing an inverse edge amounts to traversing an existing edge in the opposite direction, and depending on the data model, this may not be possible.

In a Web setting, this is prohibitively costly, because it means finding the hyperlinks pointing to a given page, which could be anywhere. Discovering all such links would require crawling the entire Web. On the other hand, distributed graph databases can be designed to efficiently support edge traversal in both directions, which is required for graph query languages such as Gremlin\footnote{http://gremlin.tinkerpop.com}. 

\subsection{Complexity}

For database query languages, the traditional parameters considered in a complexity study are the \emph{data size} and the \emph{query size}~\cite{Vardi:1982:CRQ:800070.802186}. The cost of executing a fixed query, as a function of the data size is the \emph{data complexity}, whereas if both the query and database are considered variable then the cost is the \emph{combined complexity} in Vardi's taxonomy~\cite{Vardi:1982:CRQ:800070.802186}.

The size of the data graph has two parameters, the number of nodes $|V|$ and the number of edges $|E|$.

For a regular path query, the query size is the number of characters and operators in the regular expression~\cite{Nav04jstat}, which we will note $m$.

Based on these parameters, the cost of the PAA algorithm is the cost of building the query automaton, plus the cost of building and searching the product automaton. In practice, only a subset of the product automaton will be searched and built on the fly. However, in the worst case the full product automaton is reachable and must be built and searched. 

A non-deterministic query automaton can be built in $O(m)$ time and has $O(m)$ states~\cite{Nav04jstat}. The product automaton will have $O(|V|.m)$ states and $O(|E|.m)$ transitions. 

The complexity of a graph search (BFS or DFS) over this automaton is therefore $O(|E|.m +|V|.m)$. The total combined complexity of the PAA algorithm is therefore $O((|E|+|V|).m)$. We note that a variation of RPQ where the paths should be regular \emph{simple} paths (paths where nodes are not repeated) is NP-complete~\cite{DBLP:conf/vldb/MendelzonW89}.


%


\subsection{Optimizations}

To our knowledge, two approaches for optimizing the PAA algorithm have been proposed.

The first one, introduced by Fernandez et al. in~\cite{DBLP:conf/icde/FernandezS98}, is based on graph schemas, i.e. partial knowledge of the graph structure. A graph schema can indicate that some paths, although they match a prefix of the regular expression, will not lead to a solution. These paths can then be pruned from the search space, and thus reduce the cost of the search.

The second one, by Koschmieder~\cite{DBLP:conf/gvd/Koschmieder10,DBLP:conf/ssdbm/KoschmiederL12} (Yakovets et al.~\cite{DBLP:journals/corr/YakovetsGG15} also discuss a very similar optimization), is based on knowledge of the frequency of different labels in the graph of data, and reduces the subgraph explored in the query.

If the regular expression contains a label known to be rare, the regular expression is split into smaller queries around the occurrences of the rare label, which are then used as ``waypoints" for the query execution. 

For example, for our example query $Q_2$ using the regular expression $ac(a|b)$, the label $c$ is rarer than $a$ or $b$, only occurring three times in the entire graph (whereas $a$ and $b$ occur 6 times each). The optimal query execution would then start from all edges labeled $c$, which are the edges 4-3, 2-3, and 6-8, then search backwards for $a$ (i.e., $a^{-1}$, using the notation we introduced for RPQI) and forwards for $a|b$. This produces only 3 starting points, as opposed to 6 if the search starts from every $a$ edge, and 12 if the search started from all candidates for the last edge, which may be labeled $a$ or $b$.

This approach assumes that edges can be efficiently traversed in both directions (as discussed for RPQI), and is also mainly useful for multi-source queries: for single-source queries, the starting point occurs only once, and therefore constitutes the best rare waypoint to begin the traversal from.

For both of these optimizations, it is worth noting that the basic PAA algorithm is still used. In the first case, the optimization helps prune the search of the product automaton $A_p$, while in the second case the main query is split into smaller subqueries that are also executed using the PAA. 

We can therefore consider the PAA algorithm to be the fundamental basis of RPQ processing, and in the rest of this paper we will focus on adapting this algorithm to the distributed setting, without further consideration for these optimizations.

\section{RPQ Processing on Distributed Data} 
\label{sec:distrib}

In this section, we discuss general strategies to process RPQ on distributed data.
Several query processing strategies have been proposed for SPARQL queries in general, in the Semantic Web setting (with no specific consideration of the RPQ case, which is part of the language since SPARQL 1.1), and others have been proposed for the Web and for distributed database systems. 

We discuss their applicability to the case of RPQ on non-localized data, with an evaluation of their worst-case complexity.

\subsection{Distributed Query Execution}

The main distributed query processing strategy is based on the idea of ``query shipping". In this strategy, first detailed in~\cite{abiteboul1997regular}, during the graph traversal the query is shipped to the nodes hosting the edges being traversed. This assumes that data is localized, in the sense that edges point to nodes located at a specific site of the distributed system (a specific server on the semantic web, or a specific peer in a peer-to-peer database system). There is therefore a distinction between \emph{local} edges, which point to nodes on the same server (the server where the query is being processed), and \emph{outgoing} edges, which point to a different server.

We can illustrate this on example graph of data, shown in figure \ref{fig:exgd-DS2}.
For example, to process a query $Q1 = (1, a^*bb)$, the query is first shipped to the location where node 1 is stored. Local edges matching the query are then traversed until solutions or outgoing edges are found. In our example, the local edge 1-2 (labeled $a$) is traversed, followed by the outgoing edge 2-6, also labeled $a$: the subquery $(6, a*bb)$ must be shipped to the location where node 6 is stored. The regular expression $a^*bb$ represents every possible suffix to the path $aa$, so that the full path will still match the original query. 

This approach was designed for localized data (e.g. the graph of Web pages and hyperlinks), where the node identifier would encode a single location. If the data is not localized, the location of node 6 is unknown \emph{a priori}: one solution to find it is to check every possible location, or \emph{broadcast} a query for it. The other problem is that there may be \emph{several} locations for node 6, and ideally the query should not be processed in parallel at all these sites, and instead a single location should be chosen.

We note that while the technique presented in~\cite{abiteboul1997regular} is designed for single-source queries, a similar idea is used in~\cite{DBLP:journals/tcs/ShoaranT09} for an extension to multi-source RPQ for weighted labeled graphs, where the path with minimal weight is also computed.

It is also worth noting that with localized data, supporting the ``inverse" operator is problematic, since incoming edges to a node can be located anywhere. 

\subsection{Query decomposition}

As the previous strategy may cause a large amount of traffic in shipping queries back and forth between sites, 
a better solution may be the query decomposition technique proposed by Suciu~\cite{DBLP:conf/vldb/Suciu96}, which allows a query to be answered by exchanging only one pair of messages (subquery-answer) between the querying site and each other site. 
 
The idea of this technique is to anticipate the query shipping by enumerating all the possible ``suffix" subqueries, and applying the subqueries to all ``incoming nodes", i.e. nodes that have incoming edges from other sites. The results of all these subqueries can then be collected at the querying site to reconstruct the sub-graph traversed in the query execution. 

However, this strategy again requires distinguishing ``local" and ``outgoing" edges in the graph of data, i.e. it assumes localized data.

\subsection{SPARQL Query Processing Approaches}

In the Semantic Web context, SPARQL endpoints are primarily intended to serve complex queries over their local data, acting as servers for clients which merely send a query and obtain the results. Recently, there have been a number of proposals for \emph{client-side} query processing\cite{hartigsparql2009,Harth:2010:DSO:1772690.1772733,umbrichLinkedDataQuery}: the idea is to unburden the servers (which have availability issues) and instead process the query on the client side, using data retrieved on a per-query basis, usually by dereferencing URIs (and retrieving flat RDF files).

Research in this area has largely focused on the identification of relevant data sources, and the incorporation of retrieved data into the query execution process.  

A basic approach to data source selection is to dereference the URIs found during the query processing \cite{hartigsparql2009}, which is convenient as it does not require any prior knowledge of sources. However, it implicitly relies on the assumption that the data describing a resource identified by a given URI will be found by dereferencing this URI. In other words, the assumption is that Semantic Web data is localized, and since this assumption is only partially true, techniques relying on it must accept some level of incompleteness in their results~\cite{umbrichLinkedDataQuery}.

When many sources are known, and it becomes desirable to only query a subset of these sources, more advanced query selection can make use of various types of catalogs, indexes, or summaries of the data available at each source~\cite{DBLP:journals/corr/RakhmawatiUKHH13,Harth:2010:DSO:1772690.1772733}. While these indexes and summaries can help mitigate the problem of incompleteness~\cite{Harth:2010:DSO:1772690.1772733}, in practice they are often unavailable~\cite{DBLP:journals/corr/RakhmawatiUKHH13}. 

For the actual execution of queries, query planning techniques can be adapted from the domain of relational databases, and the key issue is the ordering of JOIN operations, and of their data retrieval steps. However, these query planning techniques are primarily designed for queries with a finite number of JOINs. However, for RPQ, a path defined with a Kleene closure (*) may involve an arbitrary number of edge traversals, which are essentially JOINs, if the data is represented by a table listing the edges of the graph (e.g. RDF triples). For single-source queries, the obvious solution is then to start from the given start node, and compute the JOINs iteratively. 

Implicitly, the data for each JOIN is then also retrieved iteratively, and may involve as many iterations as nodes are traversed in the graph of data. Intuitively, this technique may retrieve only the data needed for the query, but it requires many separate requests.

This technique has been described as ``bottom-up"~\cite{Ladwig:2010:LDQ:1940281.1940311}, and can be opposed to a ``top-down" approach, which consists instead in retrieving all the data relevant to the query in a single first step, before executing the query locally. The advantage of the ``top-down" approach is that it requires a single query to each source, but the disadvantage is that in that initial step, it is much more difficult to pinpoint which data will actually be needed for the query. 

For RPQ, a simple selection strategy can use the labels that appear in the query. If the regular expression in the query doesn't contain any wildcards, only the edges with labels appearing in the query need to be retrieved. 
In our example using the query $Q1$, defined by the regular expression $a^*bb$, applying this strategy would consist in retrieving all of the edges labeled $a$ or $b$. In cases where every possible edge label appears in the query, or if the query contains a wildcard symbol (``."), this approach will result in the full graph of data being retrieved.

\subsection{Peer-to-peer Databases}
In P2P database systems (e.g. Piazza~\cite{piazza}, Edutella~\cite{Nejdl02edutella}, or SWAP~\cite{DBLP:conf/wow/EhrigTBHSSSS03}), it is assumed that a finite number of sources exist, are known and available, and query processing typically makes use of all the data sources. The data sources are connected in a P2P network and communicate via specific messages to retrieve data and process queries.

The same general approaches could also be applied, although the primary mechanism to retrieve data is then to \emph{broadcast} simple queries to the participating peers, which can then return a subset of their data matching the query.

Recent proposals of Linked Data Triple Pattern Fragments servers~\cite{verborgh2014low}, designed to support client-side query processing by serving basic filtering queries, bring the Semantic Web domain much closer to the P2P domain then before, and may shift the problem of source selection towards a problem of selecting relevant data within each source. 

If a simple query can be selective enough, it becomes conceivable to broadcast it to every Linked Data source, and obtain a much more complete set of answers, made up of very little data from each source, rather than querying few sources and retrieving larger amounts of data from each.

\subsection{Complexity Comparison}
\label{sec:DistrRPQcost}
In this section, we discuss how the above query processing techniques can be adapted for RPQ on non-localized data, and compare their asymptotic complexities for this problem. 

\subsubsection{Complexity Metrics and Parameters}

Traditional cost metrics for distributed query processing are message complexity and response time. When the distribution is used to parallelize and speed up the processing, the response time is the more relevant metric; in this case however, the distributed query processing algorithms primarily aim to reduce the communication overhead of collecting all the data in a centralized location. Therefore it is more relevant here to focus on the message cost. Ultimately, the communication overhead will also strongly affect the final response time due to network congestion.

As for centralized query processing, the size of the data graph (parameters $|E|$ and $|V|$) and the size of the query ($m$) are important complexity parameters. 

In a distributed setting, the components of the system can be represented as a graph, where the components are nodes and the edges represent their communication channels. This representation naturally unifies peer-to-peer networks, where the graph represents the overlay network, and the Semantic Web setting, where the client can be represented by a component connected in a star pattern with all the relevant data sources (servers). 

The characteristics of this graph are also essential cost factors: these factors are the number of nodes $N_p$ and the number of edges $N_c$. 

As we have mentioned previously, each data resource (atomic piece of data) may be available in zero to many arbitrary locations. However, we can consider that the data \emph{replication rate} $k$ is a cost parameter: this means that on average, each data resource is found in $K$ locations, where $K=k.N_p$. (to distinguish these, we will call $K$ the replication \emph{factor}, and $k$ the replication \emph{rate}.) We note that for asymptotic cost estimations, we will consider that $K=O(N_p)$.

\subsubsection{Communication Primitives}

In order to process a query, the sites can communicate in one of two ways: $broadcast$ or $unicast$ messages. Broadcasts are messages from one site to all the other sites, and their cost is proportional to the network size, whereas unicasts are point-to-point messages between specific sites. 
For the different query processing algorithms, broadcasts will basically be used to distributed queries (or subqueries), whereas unicasts are used to return the responses to these queries.

In the absence of network topology optimizations, and under the minimal assumptions of a reliable asynchronous network~\cite{lynchDistributedAlgorithms} the message cost of broadcasting a fixed-size message is $O(N_c)$, and we consider the cost of unicasts to be constant (independent of network size). 
 
As the centralized query processing algorithm (the PAA) is of linear complexity, the network communication will be the performance bottleneck in a distributed query execution. We therefore focus our comparison on the communication costs involved in the query execution.

\subsubsection{Top-down Query Processing} 

In this strategy, which we will call S1 in the rest of the paper, the data is retrieved by a single broadcast. If the edge labels appearing in the query are used for selecting this data, then the size of this message is proportional to the size of the query ($O(m)$). In the worst case, the full graph of data is returned. As all the peers storing data will respond, this means this amount of data is multiplied by the replication factor $K$. With $K=k.N_p$, the amount of data being sent by unicast is $O(k.N_p.(|E|+|V|))$.

 \subsubsection{Bottom-Up Query Processing}
In the ``bottom-up" query processing strategy, which we will refer to as S2 in the rest of the paper, the product automaton is constructed, with $O(m.|V|)$ nodes and $O(m.|E|)$ edges. Searching this graph, depth-first or in another order, has a cost of $O(m.(|V|+|E|))$, counting one operation to read each node or edge in the graph. 

However, as the storage is not, in fact, local, this ``read" involves a broadcast search and several unicast responses. This would imply $O(m.(|V|+|E|))$ broadcasts, and a total of $O(m.k.N_p.(|V|+|E|))$ responses sent by unicast. However, a simple optimization where the graph of data is cached locally (rather than being retrieved multiple times) reduces the number of broadcasts and unicasts to $O(|V|+|E|)$ and $O(k.N_p.(|V|+|E|))$, respectively. 

 \subsubsection{Query Shipping} 
In the ``query-shipping" strategy, which we will refer to as S3, the PAA algorithm is executed, and in the worst case every edge is an outgoing edge, meaning there is a broadcast at every step of the product automaton search, with unicast responses to return the data. 

The cost is therefore similar to that of S2, except that since the broadcasts are sent by different sites, they cannot be cached, so the amounts of data being broadcast and unicast are $O(m.(|V|+|E|))$ and $O(m.k.N_p.(|V|+|E|))$, respectively.

 \subsubsection{Query Decomposition} 
We will refer to this strategy as S4.
In the worst case, every edge at every site may be an outgoing edge. This information must be broadcast, for sites to identify their ``incoming" nodes, and therefore the message cost of this step is $O(k.N_p.|E|)$. The query execution itself can be efficient, with only one broadcast (of the initial query) and one response per site. However, in the worst case, which is that of each edge being ``outgoing", the intermediate results returned by the query could amount to the full graph of data, replicated $K$ times. The amount of data to be broadcast is therefore $O(k.N_p.|E|+m)$ and the amount of data sent by unicast is $O(k.N_p.(|E|+|V|))$.

 \begin{table*}[htb]
 \centering
 \begin{tabular}{|p{0.4cm}|p{3cm}|p{4.4cm}|p{5.3cm}|}
 \hline
 s. & broadcasts & unicasts & total \\ \hline
 S1 & $O(m)$ & $O(k.N_p.(|E|+|V|))$ & $O(m.N_c + k.N_p.(|E|+|V|))$\\ \hline
 S2 &  $O(|V|+|E|)$ & $O(k.N_p.(|E|+|V|))$ & $O(k.N_p.N_c.(|E|+|V|))$\\ \hline
 S3 & $O(m.(|E|+|V|))$ & $O(m.k.N_p.(|E|+|V|))$& $O(m.N_c.(|E|+|V|)+m.k.N_p.(|E|+|V|))$ \\ \hline
 S4 & $O(k.N_p.|E|+m)$ & $O(k.N_p.(|E|+|V|))$ & $O(k.N_p.(N_c.|E|+|V|)+N_c.m)$\\ \hline
 \end{tabular} 
 \caption{RPQ for non-localized data: summary of asymptotic query processing costs.}
 \label{tab:strategycosts}
 \end{table*}

\subsection{Conclusion}

The study of worst-case complexity paints the somewhat naive ``top-down" query processing strategy S1 in a very favourable light. However, its worst case still involves transferring the entire graph of data over the network, which is problematic. In addition, this worst case appears quite likely to happen: with our simple selection method, the mere presence of a wild-card in the query is enough to cause it. 

In any case, it appears that with every strategy, we have an inescapable worst case where computing a query actually requires the full graph of data, and therefore the message cost can reach at least $\Theta(|E|+|V|)$.

With very large datasets, even this baseline complexity seems unmanageable. This motivates a search for alternative strategies to reduce the costs: can we anticipate very costly query processing by analyzing the query? Can we cap the cost by interrupting the processing once a limit has been reached?

For the second question, it seems clear that for strategies S1, S3, and S4, the answer is no: once the query is sent out, the querying agent has little visibility or control over the processing. This is an advantage of the iterative strategy S2.

In addition, although more costly than S1 in the worst case, S2 also has the advantage that it only retrieves the data needed for the query: this implies that for \emph{selective} queries, S2 may perform better.

We can compare the amounts of data broadcast and retrieved for some ``real-world" queries. Given some concrete figures, we can then analyze the trade-off between these quantities, and determine conditions that favour one or the other strategy.
 
This is the object of the next section.

\section{Cost Comparison on Real-world Queries}
\label{sec:discriminant}
\subsection{Dataset and queries}
In order to conduct an empirical study, we acquired a ``real-world" dataset from the biomedical domain, with some meaningful queries to apply to this data. The dataset is a graph of knowledge automatically extracted from a corpus of pubmed abstracts~\cite{plake2006alibaba}. The queries are the same ones used in~\cite{DBLP:conf/ssdbm/KoschmiederL12}; the queries and the dataset were kindly provided by the authors of that study.

The graph has approximately 50,000 nodes and 340,000 edges. The nodes identify concepts such as molecules, genes, or animal species, and the edges represent relationships, such as a molecule activating a gene or the terms simply co-occurring in the same pubmed abstract. 

The queries express meaningful associations between biological entities, and are expressed as multi-source regular path queries, i.e. regular expressions on the alphabet of edge labels. These regular expressions are listed in table~\ref{tab:12queries}, with the number of solution pairs to the multi-source query. 

\begin{table*}[tp]
\centering
 \begin{tabular}{|p{0.5cm}|p{4cm}|p{3cm}|p{3.5cm}|}
 \hline
 &\textbf{Query} & \textbf{Multi-source solution pairs} & \textbf{Start nodes with non-zero cost}\\ \hline
q1&C$^+$ ``acetylation" A$^+$ & 1710 & 477 \\ \hline
q2&C$^+$ ``acetylation" I$^+$ & 20 & 477 \\ \hline
q3&C$^+$ ``methylation" A$^+$ & 0 & 477 \\ \hline
q4& C$^+$ ``methylation" I$^+$ & 0 & 477 \\ \hline
q5& C$^+$ ``fusions" P & 0 & 477 \\ \hline
q6& ``fusions" A$^+$ & 8 & 2\\ \hline 
q7& A$^+$ ``receptor" P & 0&731 \\ \hline
q8& I$^+$ ``receptor" P & 0&366  \\ \hline
q9& A A$^+$ & 80905 & 711\\ \hline
q10& I I$^+$ & 2118 & 354\\ \hline
q11& C E & 249 & 364\\ \hline
q12& A$^+$ I$^+$ & 49638 & 711\\ \hline
 \end{tabular} 
\\
~\\
\textbf{Edge labels} (`\textbar' means disjunction): \\
\begin{tabular}{p{1cm}p{15cm}}
\hline

C =&\small{\{interaction \textbar ~interactions \textbar ~binding \textbar ~complex \textbar ~interacting \textbar ~complexes \textbar ~interacts\}}\\
A =&\small{\{activation \textbar ~activity \textbar ~production \textbar ~induction \textbar ~overexpression \textbar ~up-regulation \textbar ~induces \textbar ~activates \textbar ~increases\}}\\
I =&\small{\{down-regulation \textbar ~inhibits \textbar ~inhibited \textbar ~inhibitor \textbar ~inhibition\}}\\
E =&\small{\{expression \textbar ~overexpression \textbar ~regulates \textbar ~up-regulation \textbar ~expressing\}}\\
P =&\small{\{dephosphorylates \textbar ~dephosphorylated \textbar ~dephosphorylate \textbar ~dephosphorylation \textbar ~phosphorylates \textbar ~phosphorylated \textbar ~phosphorylate \textbar ~phosphorylation\}}\\
\hline
\end{tabular}
 \caption{Biological queries: regular expression, number of solutions (multi-source), number of valid start nodes}
 \label{tab:12queries}
 \end{table*}

As the graph has 50,000 nodes, we can create 50,000 single-source queries from each regular expression (multi-source query). However, for many nodes the relationship expressed by the query simply does not make sense, which in practice means for most nodes there will be no adjacent edges matching the beginning of a query path, and therefore the cost of evaluating the query will be basically nil. For the queries of interest here, less than 2\% of the nodes were valid starting points. The exact number of valid starting points for each query is given in the last column of table~\ref{tab:12queries}.

When evaluating the cost of queries, we will restrict our analysis to these valid starting points. Obviously, calculating the average cost when it is nil 99\% of the time would produce results of little value.

In order to compare the cost of strategies S1 and S2 for these queries, one way would be to evaluate them in a distributed setting, and vary the parameters of this distributed setting. However, for these strategies, the distribution parameters only directly affect the cost of broadcasts or unicasts, which in turn determine the cost of the query execution. We can therefore compute the number of broadcasts and unicasts required for each query, then calculate the costs of each broadcast and unicast analytically for the different values of the distribution parameters.

For this purpose, we first compute the number of unicasts and broadcasts for these different queries. 

\subsection{Broadcast and Unicast costs}
In the following analysis, we consider each symbol (edge label or node identifier) transmitted in a query or response as the unit of cost for message traffic. The true cost would be obtained by multiplying our values by the number of bytes per symbol, and adding some overhead for the message headers. 

\subsubsection{Cost of S1}
The processing of a query by S1 requires a single initial query (broadcast) to retrieve all the data (a subset of $G_D$) potentially needed to process the RPQ. 

This data is the set of edges whose labels are found in the query, and the length of the initial broadcast query is therefore the number of distinct labels in the query. 

The data matching the query is then returned by ``point-to-point" messages (unicasts). The amount of data to be transferred is the number of matching edges to the initial broadcast, multiplied by the length of each edge: we consider that an edge is expressed as 3 symbols, two node identifiers and an edge label.

It is important to notice that for strategy S1, the cost does not depend on the query start node, and is even the same for a single-source or a multi-source query.

\subsubsection{Cost of S2}

In strategy S2, the PAA is executed locally, accessing the remote data through broadcast searches. 

During the search of the product automaton, at each node there is a broadcast search to find neighbours of the current graph node. Each time, the broadcast query indicates the current node and the labels of the potential outgoing edges, which are the symbols associated with the outgoing transitions from this automaton state.

The amount of data to be broadcast is the sum of the lengths of all these individual queries.
The data returned as unicast messages is the set of edges in $G_D$ that match them. 

We assume a simple optimization whereby two identical broadcast queries that are made at different points of the algorithm will result in a single one being processed over the network, and the second time its results are obtained from a local cache.

Unlike strategy S1, for strategy S2 each single-source query will have a different cost.
\subsection{Results}

\begin{figure*}[t]
\centering
\begin{subfigure}[b]{0.50\textwidth}
 \includegraphics[scale=0.31]{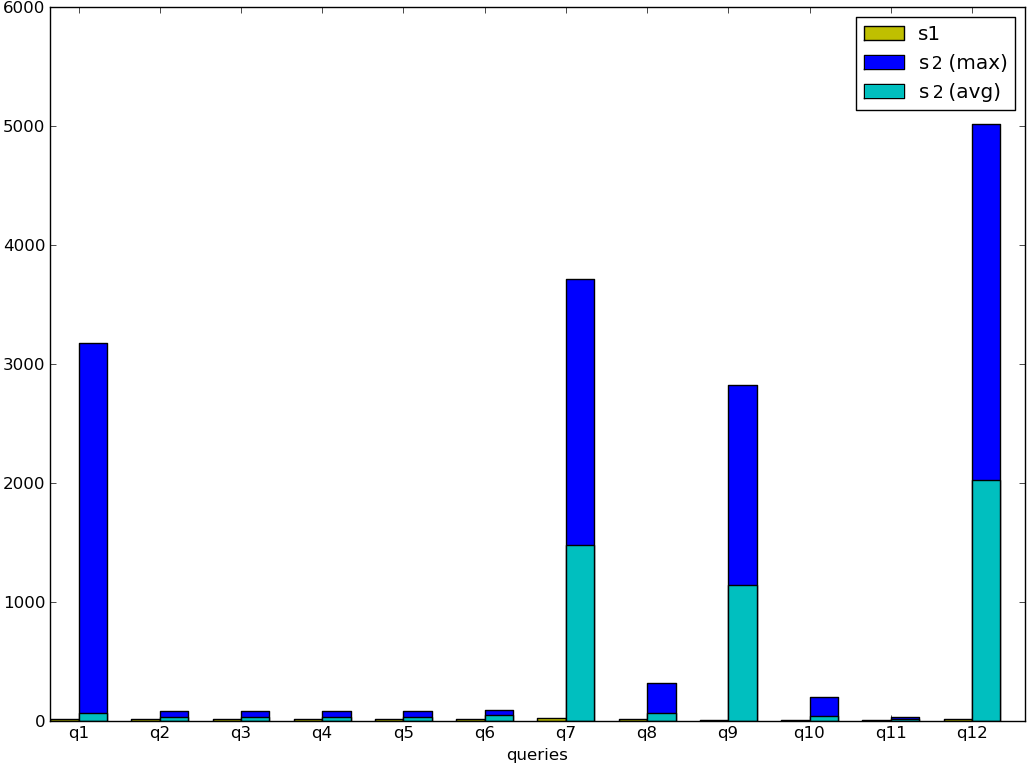}
\caption{Broadcasts}
\label{fig:broadcasts}
\end{subfigure}
\begin{subfigure}[b]{0.49\textwidth}
 \includegraphics[scale=0.32]{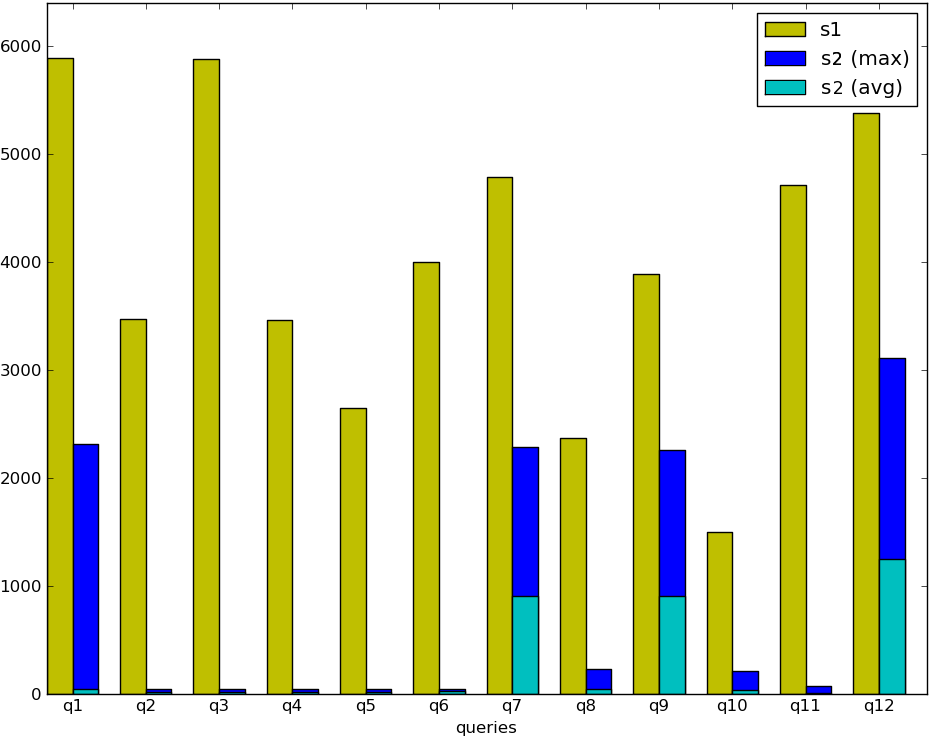}
\caption{Unicasts}
\label{fig:unicasts}
\end{subfigure}
\caption{Amount of data to be transferred by broadcast /unicast for the evaluation of the biological queries.}
\end{figure*}

Figures \ref{fig:broadcasts} and \ref{fig:unicasts} summarize the values obtained for the above cost functions, with our biological queries.

For each regular expression, we compare side-by-side the cost of S1 (which is the same for all the single-source queries) and the costs of S2: we show here the mean and the maximum cost. Note that the mean is calculated only for valid starting points, i.e. it is the mean of all non-zero costs.

These figures illustrate well the trade-off between broadcasts and unicasts for the two strategies.
Strategy S1 always requires a minimal amount of data to be broadcast, but also consistently retrieves fairly large amounts of data via unicast. Strategy S2 has typically higher broadcast costs and much lower unicast costs, and is also much more variable. We note that the queries considered here are all quite selective, in the sense that they only retrieve a small fraction of the data graph. S1 retrieves between 0.2\% and 0.8\% of the graph, whereas S2 retrieves less than 0.1\% of the graph in almost every case.

However, due to this trade-off and the high variations in the cost of S2, it is unclear which strategy is generally preferable. In the following, we examine this trade-off in analytical terms, in relation to the parameters that determine the costs of broadcast and unicast messages.

\subsection{Cost functions}

In order to express and compare the costs of S1 and S2, we introduce the following notations:

\begin{itemize}
 \item the number of distinct labels in a query $q$ is noted $Q_{lbl}(q)$,
 \item the number of matching edges, or rather the amount of data encoding these edges, is noted $D_{S1}(q,G_D)$,
 \item $Q_{bc}(q,G_D)$ describes the total amount of data that is broadcast with strategy S2,
 \item The amount of data transferred by unicast for S2, encoding the set of edges traversed by the algorithm, is noted $D_{s2}(q,G_D)$. 
 \item The data is replicated on average $K$ times, where $K=k.N_p$. 
\end{itemize}

In the above functions we have indicated the dependencies of these quantities on $q$ and $G_D$ as function arguments. In the following we leave these function arguments out in order to improve readability. 

We also remind the reader that the number of messages for a broadcast in a connected network with $N_c$ edges is between $N_c$ (best case) and $2.N_c$ (worst case). If the average (outgoing) node degree in the network graph is $d$, then $N_c$ can be approximated as $d.N_p$. Ignoring the protocol overhead for each message, we can therefore approximate the cost of broadcasting $b$ bytes of data as $2.d.N_p.b$. 

We obtain the following cost functions:
\begin{align}
\label{eq:costs1rdf}
 cost_{S1}(q,G_D)	& = 2.N_c.Q_{lbl} + k.N_p.D_{s1} \notag \\
			& = 2.d.N_p.Q_{lbl} + k.N_p.D_{s1} \notag \\
			& = N_p(2.d.Q_{lbl} + k.D_{s1})
\end{align}
\begin{align}
\label{eq:costs4rdf}
cost_{S2}(q,G_D)	& = 2.N_c.Q_{bc} + k.N_p.D_{s2} \notag \\
			& = N_p(2.d.Q_{bc} + k.D_{s2})
\end{align}
\subsection{Query Execution Strategy Choice}
\label{sec:discriminantconditions}
Using equations \ref{eq:costs1rdf} and \ref{eq:costs4rdf}, we can establish the following condition, determining whether S2 or S1 is preferable :
\begin{align} 
\label{eq:discr}
~ & ~ 2.d.Q_{lbl}+k.D_{s1} < 2.d.Q_{bc} + k D_{s4} \notag \\
\Leftrightarrow & ~ k.(D_{s1} - D_{s2}) < 2.d (Q_{bc} - Q_{lbl}) \notag \\
\Leftrightarrow & ~ \frac{k}{d} < 2 \frac{Q_{bc} - Q_{lbl}}{D_{s1} - D_{s2}}
\end{align}
 In the following, we will use the notation:
\begin{equation*}
 discr(q,G_D) = 2\frac{Q_{bc} - Q_{lbl}}{D_{s1} - D_{s2}} 
\end{equation*}

Equation \ref{eq:discr} 
provides a discriminating condition to choose between S1 and S2, independent of the network size. Parameters $d$ and $k$ characterize the network topology and the data distribution within this network. Higher values of $d$ (denser networks) increase the cost of broadcasts, therefore favouring strategy S1, that broadcasts less data, whereas higher values of $k$ (higher data replication rates) increase the cost of retrieving data (each data resource retrieved comes in more copies), therefore favouring S2, which only retrieves the data necessary to execute the PAA.

We also know that $k<1<d$, because $k>1$ would mean that every peer has multiple copies of the full graph of data, and in a network where $d<1$ the network graph cannot be connected\footnote{with the exception of the linear alignment of $n$ peers, where $d=\frac{n-1}{n}$}. 

\begin{figure}[htb]
\centering
 \includegraphics[scale=0.5]{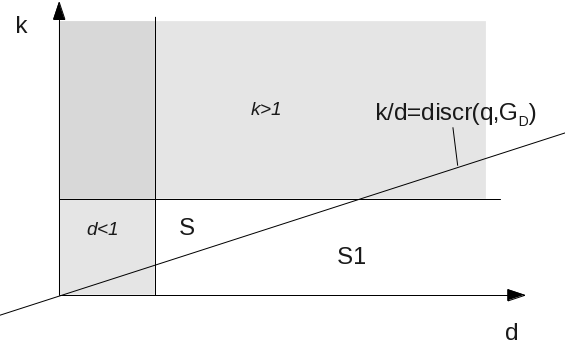}
\caption{Optimality of S1 and S2 depending on $k$, $d$ and the query-dependent discriminating function.}
\label{fig:discr}
\end{figure}

The sets of values for which S2 and S1 are optimal can be visualized graphically as in figure \ref{fig:discr}.


This gives us the following discriminating conditions:
\begin{itemize}
 \item If $Q_{bc}(q,G_D) \leq Q_{lbl}(q,G_D)$ then S2 is necessarily optimal. The trivial case is where the query starting point is not valid, and this may also happen with very long and complex queries.
 \item If $Q_{bc}(q,G_D) > Q_{lbl}(q,G_D)$, then in the 2-dimensional space of values for $k$ and $d$, S2 is optimal in a triangle bounded by the lines of equations $k=1$, $d=1$ and $\frac{k}{d} = discr(q,G_D)$.
 \item For any other values of $k$ and $d$ that fulfil the condition $k<1<d$, S1 is optimal.
 \item Note that if $discr(q,G_D)>1$, then S1 is necessarily optimal, because the triangle described above does not intersect with the region where $k<1<d$.
\end{itemize}

For our example biomedical queries, of the 5622 single-source queries with non-zero cost\footnote{Altogether we could apply each of the 12 multi-source queries to 50,000 nodes, yielding 600,000 single-source queries.}), in 42 cases S2 is necessarily optimal, and for the 5580 others, either S1 or S2 will be optimal depending on the network parameters.

The problem is that while we can immediately identify queries with zero cost, and have immediate access to $Q_{lbl}(q)$ (the number of distinct labels in the query), the other parameters are not readily accessible. Therefore, $discr(q,G_D)$ cannot be calculated without actually executing the query over the data of interest, which means that despite this analysis we still cannot tell, \emph{a priori}, which strategy to choose.

One solution would be to estimate the remaining cost parameters, which may be possible if we have access locally to a sample of the data. Such an estimate would have the added benefit that we could detect the ``worst case" situations, where the full graph of data is likely to be downloaded, causing an overload of network traffic.

The next section will explore two statistical graph models that can be used to produce such cost estimations.
\section{Query Cost Estimation}
\label{sec:costeval}

\subsection{Related Work}

To the best of our knowledge, the only existing method to estimate query selectivity is the ``locality" metric proposed in~\cite{DBLP:journals/jodl/MendelzonMM97}. This metric mainly relies on the distinction between ``local" and ``outgoing" edges in the query itself to determine how many sites need to be visited to answer the query, and assumes that a bounded amount of data is stored at each site. In most query languages (and in the basic definition of RPQ) this distinction between local and outgoing edges does not exist.

Several data structures have also been proposed to ``summarize" graph data, which allows a user to view the structure of the data and the existing paths to be queried, and can also be used to optimize query execution by pruning the search graph~\cite{ilprints264,DBLP:conf/icde/FernandezS98,Chen:2003:DIA:872757.872776}. These techniques mainly give indications of whether paths exist, with no direct applications to cost estimation.

For XML data, some techniques to estimate the selectiveness of path queries were proposed in the context of the XPath query language~\cite{Lim:2002:XOS:1287369.1287408}. These techniques were designed for tree-structured data and do not handle regular expressions. However, our approach is comparable, extended to arbitrary graphs and regular paths.

\subsection{Estimation}
\subsubsection{Network and Distribution Parameters}
\label{sec:networkParamEstimation}
These parameters will determine the cost of a single broadcast, and the multiplying factor for responses returned by multiple peers. These parameters can be obtained or estimated from simple queries. Ideally, these queries should be supported by the communication protocol.

\begin{itemize}
 \item $N_p$, network size: can be obtained by broadcasting a ``ping" query, where each recipient simply responds with an acknowledgement. 
 \item $N_c$, number of network connections: can be obtained by broadcasting a query requesting each peer's number of active network connections. The sum of responses will be $2*N_c$.
 \item $k$, data replication rate: can be estimated by querying for a small number of known data resources, and counting the average number of responses. This will yield an estimate of $K$ (data replication factor), which can then be divided by $N_p$ to obtain $k$. The accuracy of this estimate depends on using a representative sample of resources. Querying for more resources incurs more costs but will improve the estimate.
\end{itemize}

\subsubsection{Query selectivity}

We now turn to the ``query selectivity" functions:

\begin{itemize}
 \item $Q_{lbl}(q)$ can be trivially obtained from the query: it is the number of distinct edge labels appearing in the regular expression. 
 \item $D_{s1}(q,G_D)$ can be estimated by counting the label frequencies on a sample of the data, and multiplying by the total number of edges $|E|$. $|E|$ can be estimated by a broadcast query requesting a count of the distinct resources stored at each site, then dividing by the expected replication $K$.
 \item $Q_{bc}(q,G_D)$ and $D_{s2}(q,G_D)$ are more difficult to estimate. The message cost of executing a query is linear in the data size, but only because in the worst case it requires transferring exactly the full data graph. In terms of \emph{true cost}, there is no evidence of a linear relationship between the cost on a sample of the data and the true cost, if only because it is highly unlikely that the start node will be present in the sample. Instead, we have explored the possibility of creating statistical models of the graph of data, then evaluating the query against a larger random graph generated from this model. We have considered two models, which we describe next.

\end{itemize}

\subsection{Statistical Graph Models}

\subsubsection{Binomial Random Graph}
The first model that we investigate is based on the binomial random graph model (or Gilbert model~\cite{gilbert1959}), where for any pair of nodes $(v_1, v_2)$, there is a probability $p$ that the edge $(v_1, v_2)$ exists. Extending this model to labelled graphs, for any label $a$, each edge $(v_1, a, v_2)$ exists with a probability $p(a)$.

The probabilities $p(a_i)$ for the different edge labels can be estimated by frequency counts.

Based on this model, the cost of a query can be estimated by executing the PAA replacing the access to the data graph with a function that randomly generates edges using the binomial distribution.

\subsubsection{Bayesian-Binomial Random Graph}
The disadvantage of the above model is that it completely ignores the graph structure, in the sense that adjacent edges have independent probabilities of existing. In fact, in a real-world graph, it is likely that the presence and labels of adjacent nodes are correlated, due to the semantics of the relationships that they represent. 

A more elaborate model should therefore estimate the probabilities of edges conditional to the existence (and labels) of adjacent edges. Although such a ``static" model is difficult to describe, we can use a ``generative" model: as above we apply the PAA and replace the access to the data graph with a function that randomly generates edges, except this time we generate edges using probabilities conditional to the label of the edge that brought us to this node.

\subsection{Evaluation}

Based on the above models, we have evaluated the quality of our cost evaluations using our biological queries.

As we have noted previously, for S2 the cost factors (the amount of data that is broadcast, and the amount of data retrieved by unicast) vary wildly. However, they are also highly correlated, and we will focus here on results for the number of edges traversed during query execution, which can be considered an (inverse) measure of the query selectivity. It seems unrealistic to hope to estimate the cost of a specific single-source query without a detailed knowledge of the graph of data, since applying the same path query to different (valid) starting points yields very different costs. However, these costs follow a specific probability distribution, which we can consider to be a good target for estimation.

For each query, we have therefore compared the real distribution (frequencies) of costs for the different start nodes, with the distribution of costs obtained for many runs of our algorithm, using each of the models. Specifically, since the graph has 50,000 nodes, we have 50,000 true costs, and we also ran 50,000 runs of the algorithms for the models. Recall that 99\% of the time, the cost is nil; this was true for the models as well, within one or two percent.

We constructed the models using the full graph of data: this allows us to evaluate the technique itself, without the noise caused by imperfect sampling\footnote{Ideally, a \emph{representative} sample of the data should produce the same graph label frequencies, and therefore the same models. Variations due to imperfect sampling are not our main concern here.}. We defer to future work the evaluation of approximate models obtained using only a sample of the data. 

\begin{figure}[tp]
\centering
\begin{subfigure}[b]{0.49\textwidth}
 \includegraphics[scale=0.41]{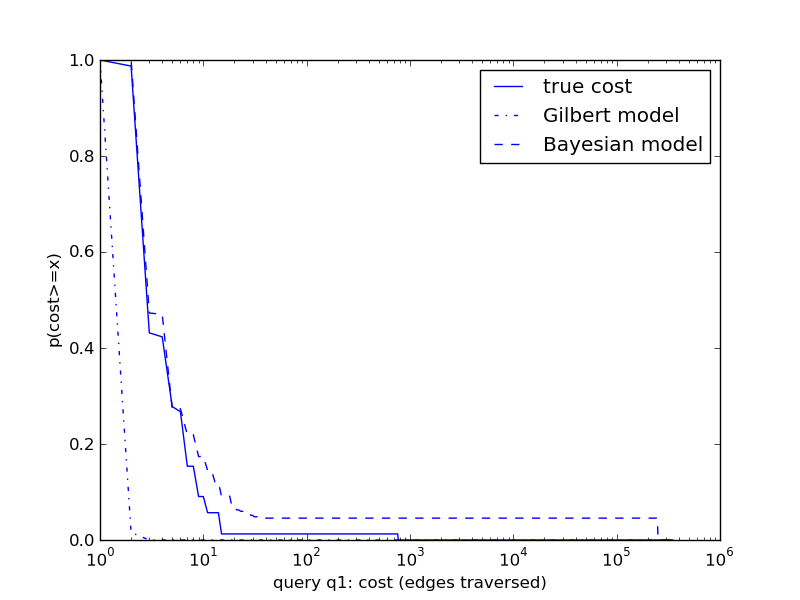}
\caption{Tail Distribution for query q1.}
\label{fig:tailsd-q1}
\end{subfigure}%

\begin{subfigure}[b]{0.49\textwidth}
 \includegraphics[scale=0.41]{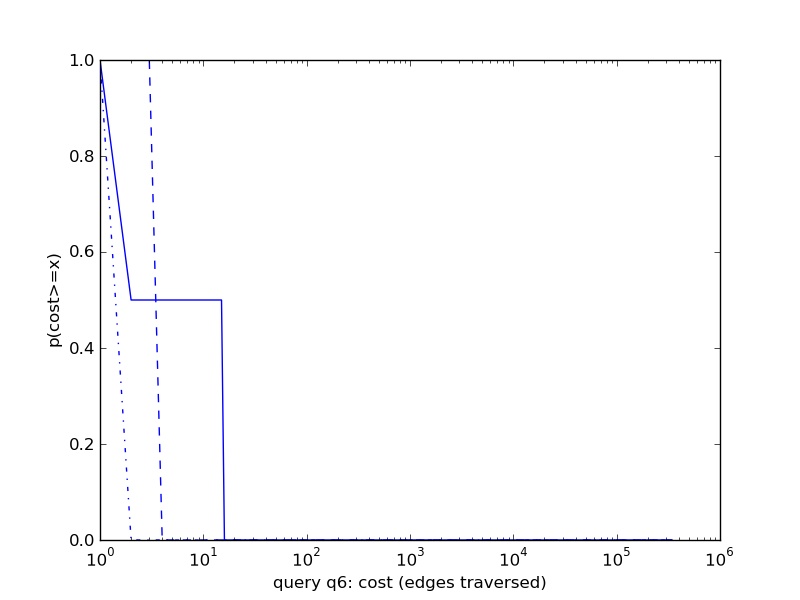}
\caption{Tail Distribution for query q6.}
\label{fig:tailsd-q6}
\end{subfigure}

\begin{subfigure}[b]{0.49\textwidth}
 \includegraphics[scale=0.42]{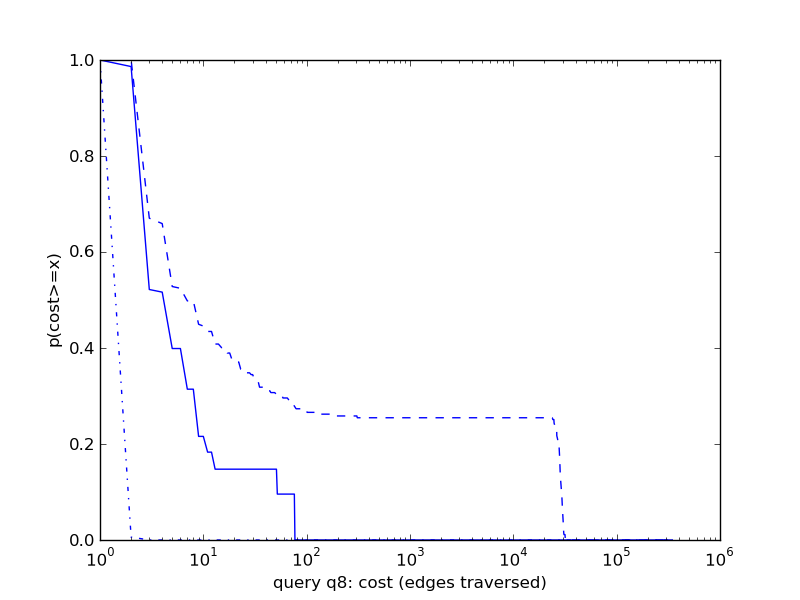}
\caption{Tail Distribution for query q8.}
\label{fig:tailsd-q8}
\end{subfigure}

\caption{Tail Distributions for different queries: true distribution, and estimates based on the two types of statistical graph models.}
\label{fig:taildistr}
\end{figure}

In order to compare the probability distributions of the true costs and the models, we have plotted the three tail distributions (complementary CDF) for each query: figure~\ref{fig:taildistr} shows the plots for a sample of the queries (4 queries). The others are comparable. Note the logarithmic scale on the x-axis.

Across all queries, we observe that the first model consistently underestimates costs, whereas the Bayesian model tends to overestimate them, although there are exceptions (as for query 6, shown in figure~\ref{fig:tailsd-q6}).

As we have pointed out previously, there are no established metrics or baselines to evaluate these models, so we will limit ourselves to an informal explanation of these resulting figures.

The reason why the simple Gilbert model underestimates the likelihood of paths is that it ignores the structure of the graph and therefore its semantics. As we have mentioned before, for a given edge label, there are only very few nodes for which the relationship expressed by this label makes sense. This could mean that at any point in a path, there is a very low probability that such a path continues. And this is the result obtained by the model: paths are very short. However, as the query (and the data) \emph{is} meaningful, at any point along a path, edges will tend to lead to nodes for which the relationship \emph{does} make sense, and the probability that the path continues is higher than that found by the Gilbert model.

This is precisely the point of using conditional probabilities in the Bayesian model. The Bayesian model estimates the probabilities of outgoing edges of a node conditionally to the label by which we reached this node. 

For example, we could be following paths labelled $a^*bb$, where the label $b$ might be very rare, but such labels may be clustered together due to the semantics of the relationship $b$. This would mean that the probability of an edge $b$ existing between two arbitrary nodes $v_1$ and $v_2$ may be very low, but might be much higher if we know that $v_1$ has an incoming edge labelled $b$. 

 Where the Bayesian model falls short of perfection is that although it may produce good estimates of the number of outgoing edges from a given node, it then picks the targets of these edges at random, ignoring other structural properties of the graph such as \emph{clustering} (in an undirected graph) or edge \emph{transitivity}, the equivalent in a directed graph. These properties mean that in real-world graphs, if two nodes $v_1$ and $v_2$ have a common neighbour $v_3$, they are more likely to be themselves connected than would be expected by random chance. This implies that paths with a common origin will tend to merge together and explore fewer nodes (than would be expected in a random graph without those structural properties). 

\section{Example Scenario}
\label{sec:scenario}
Before concluding the chapter, as a way of illustrating the importance of the different analysis and techniques developed here, we revisit the problem of choosing a query execution strategy, with an example scenario.

We consider a biomedical researcher (we will name her Alice) involved in a data-sharing network. In this network, there are 150 researchers, with an average of 6 connections each. The data they share is the Alibaba dataset (the dataset we have been using throughout this chapter), and each data item (edge in the graph) is shared by 20\% of the researchers, on average.

Alice is looking for enzymes linked to the protein known as $p53$, of central importance in cancer research. 

Specifically, Alice would like to apply the single-source query $q = (p53, C^+ \text{``acetylation"} A^+)$ to the shared dataset.

The regular expression is that of query $1$ shown in table \ref{tab:12queries}, and $C$ and $A$ refer to groups of labels given in the same table. $C$ is a list of different terms describing protein interaction, and $A$ is a list of terms describing the activation of chemical compounds (e.g. enzymes). 

In this scenario, the main questions are: how should Alice evaluate the query in a way that minimizes the network traffic?  How much traffic might the query generate?

Following the analysis in section \ref{sec:discriminantconditions}, we use the discriminating function:
\begin{equation*}
 discr(q,G_D) = 2\frac{Q_{bc} - Q_{lbl}}{D_{s1} - D_{s2}}
\end{equation*}

Alice can now estimate the different components of this discriminant function:
\begin{itemize}
\item  The network and distribution parameters $N_p$, $N_c$ and $k$ are obtained as described in section \ref{sec:networkParamEstimation}. Alice obtains the values 150, 450, and 0.2. In addition, she estimates the network density $d$ to be $\frac{N_c}{N_p}= 3$.
 \item There are 18 distinct edge labels in this query, so $Q_{lbl}=18$
\item Alice estimates $D_{s1}$ on her local data: the labels in the query represent approximately 0.5\% of the edges in the data. This means over the total graph she can expect a value of around 1800 for $D_{s1}$.
\item We assume that Alice's data produces accurate edge label frequencies, and her estimate of the distributions for $D_{s2}$ are as shown in figure \ref{fig:tailsd-q1}. We note that in this model, the distributions are given only for non-zero costs: Alice knows that the protein $p53$ is known to interact with many other proteins, so she is certain that there are edges labelled $A$ adjacent to the start node; therefore the cost will not be zero.
 According to the model, she estimates that there is 90\% chance that $D_{s2}$ will be less than 15, and around 10\% chance that it will be much higher. However, despite the very high values generated by the Bayesian model, she knows that it is bounded by the value of $D_{s1}$ (1800). 
\item Alice's model for broadcast costs (not shown) provides her with a similar-looking figure, and she finds there is a 90\% chance that $Q_{bc}$ will be less than 70, and a 10\% chance that it will reach a higher value. Using\footnote{There is an approximately linear relationship between $D_{s2}$ and $Q_{bc}$, we omit the details here.} the bounds for $D_{s2}$, Alice estimates that this higher value will be of the order of 8000.
\end{itemize}

For the higher values of $Q_{bc}$ and $D_{s2}$, the use of the discriminant function is unnecessary: it is obvious that the unicast costs of the two strategies will be comparable, whereas S2 will have a very high broadcast cost.

For the lower estimates of $Q_{bc}$ and $D_{s2}$, the discriminant function is:
\begin{align}
 discr_{low} &= 2\frac{70 - 18}{1800 - 15} \notag \\ 
 &= 0.058    \notag
\end{align}
and: 
\begin{align}
 \frac{k}{d} &= \frac{0.2}{3} \notag \\ 
 &= 0.067    \notag
\end{align}

Therefore, $\frac{k}{d}>discr_{low}$. This means that according to these estimates, S2 has a 90\% chance of being better, whereas there is a 10\% chance that S1 is better.

Alice can now make an informed choice, based on estimates that required mainly local processing, and a small number of broadcast queries (to estimate the network size and data replication rate, as discussed in section \ref{sec:networkParamEstimation}).

It is worth noting that even if S1 turns out to be less costly, with S2 Alice can also set an arbitrary limit to the cost, and interrupt the query if the limit is reached. This, of course, would come at the expense of completeness.



\section{Conclusion}
 \label{sec:conclusion}
We studied the problem of evaluating regular path queries on distributed data, where data is not localized, and is accessed through broadcast or unicast messages.

Our complexity analysis shows that the worst-case message cost of all the known algorithms is equivalent to collecting the full graph of data locally (or higher), which can be prohibitively expensive. 

We compared the most promising approaches, which are the centralized ``top-down" and ``bottom-up" approaches (S1 and S2, respectively). Our analytical and experimental comparison of these two approaches showed that S1 generates most of its cost by retrieving more data than necessary, while S2 only retrieves the data that it needs, but requires more broadcasts to locate this data during its execution.

Therefore, S2 performs better on more selective queries, where S1 is very wasteful. However, until now there were no known techniques to estimate the selectivity of a query without executing it on the dataset of interest. In order to address this problem, we proposed a query cost estimation approach based on two classes of statistical graph models. This ultimately provides a way to choose between S1 and S2, and to evaluate the cost of processing a query before potentially flooding the network with an excessive amount of traffic.

The above scenario shows how our techniques can help support decisions regarding query execution strategies. It is important to note that we have considered broadcasts to be queries distributed to all the data sources: however, these results generalize to situations where alternative techniques (e.g. an index) can be used to locate data items. In such cases, the discriminant function has to be modified to account for the new broadcast cost, but the rest of our approach is still applicable. With a lower broadcast cost, strategy S2 becomes more attractive.


\bibliographystyle{abbrv}
\bibliography{rwj}

\begin{thebibliography}{10}

\bibitem{abiteboulvianuqueriesandComputationWeb}
S.~Abiteboul and V.~Vianu.
\newblock Queries and computation on the web.
\newblock In F.~Afrati and P.~Kolaitis, editors, {\em Database Theory — ICDT
  '97}, volume 1186 of {\em Lecture Notes in Computer Science}, pages 262--275.
  Springer Berlin Heidelberg, 1997.

\bibitem{abiteboul1997regular}
S.~Abiteboul and V.~Vianu.
\newblock Regular path queries with constraints.
\newblock In {\em Proceedings of the sixteenth ACM SIGACT-SIGMOD-SIGART
  symposium on Principles of database systems}, pages 122--133. ACM, 1997.

\bibitem{DBLP:conf/pods/Baeza13}
P.~Barcel{\'o}~Baeza.
\newblock Querying graph databases.
\newblock In R.~Hull and W.~Fan, editors, {\em PODS}, pages 175--188. ACM,
  2013.

\bibitem{DBLP:journals/sigmod/CalvaneseGLV03}
D.~Calvanese, G.~{De Giacomo}, M.~Lenzerini, and M.~Y. Vardi.
\newblock Reasoning on regular path queries.
\newblock {\em {SIGMOD} Record}, 32(4):83--92, 2003.

\bibitem{Chen:2003:DIA:872757.872776}
Q.~Chen, A.~Lim, and K.~W. Ong.
\newblock D(k)-index: An adaptive structural summary for graph-structured data.
\newblock In {\em Proceedings of the 2003 ACM SIGMOD International Conference
  on Management of Data}, SIGMOD '03, pages 134--144, New York, NY, USA, 2003.
  ACM.

\bibitem{Consens:1990:GVQ:93597.98748}
M.~P. Consens and A.~O. Mendelzon.
\newblock The {G}+/{G}raphlog visual query system.
\newblock In {\em Proceedings of the 1990 ACM SIGMOD International Conference
  on Management of Data}, SIGMOD '90, pages 388--, New York, NY, USA, 1990.
  ACM.

\bibitem{cruz1987graphical}
I.~F. Cruz, A.~O. Mendelzon, and P.~T. Wood.
\newblock A graphical query language supporting recursion.
\newblock In {\em ACM SIGMOD Record}, volume~16, pages 323--330. ACM, 1987.

\bibitem{DBLP:conf/wow/EhrigTBHSSSS03}
M.~Ehrig, C.~Tempich, J.~Broekstra, F.~van Harmelen, M.~Sabou, R.~Siebes,
  S.~Staab, and H.~Stuckenschmidt.
\newblock Swap - ontology-based knowledge management with peer-to-peer
  technology.
\newblock In Y.~Sure and H.-P. Schnurr, editors, {\em WOW}, volume~68 of {\em
  CEUR Workshop Proceedings}. CEUR-WS.org, 2003.

\bibitem{DBLP:conf/icde/FernandezS98}
M.~F. Fernandez and D.~Suciu.
\newblock Optimizing regular path expressions using graph schemas.
\newblock In S.~D. Urban and E.~Bertino, editors, {\em ICDE}, pages 14--23.
  IEEE Computer Society, 1998.

\bibitem{gilbert1959}
E.~N. Gilbert.
\newblock Random graphs.
\newblock {\em The Annals of Mathematical Statistics}, 30(4):1141--1144,
  December 1959.

\bibitem{ilprints264}
R.~Goldman and J.~Widom.
\newblock Dataguides: Enabling query formulation and optimization in
  semistructured databases.
\newblock Technical Report 1997-50, Stanford InfoLab, 1997.

\bibitem{piazza}
A.~Halevy, Z.~Ives, J.~Madhavan, P.~Mork, D.~Suciu, and I.~Tatarinov.
\newblock The {P}iazza peer data management system.
\newblock {\em Knowledge and Data Engineering, IEEE Transactions on},
  16(7):787--798, July 2004.

\bibitem{Harth:2010:DSO:1772690.1772733}
A.~Harth, K.~Hose, M.~Karnstedt, A.~Polleres, K.-U. Sattler, and J.~Umbrich.
\newblock Data summaries for on-demand queries over linked data.
\newblock In {\em Proceedings of the 19th International Conference on World
  Wide Web}, WWW '10, pages 411--420, New York, NY, USA, 2010. ACM.

\bibitem{hartigsparql2009}
O.~Hartig, C.~Bizer, and J.-C. Freytag.
\newblock Executing sparql queries over the web of linked data.
\newblock In A.~Bernstein, D.~Karger, T.~Heath, L.~Feigenbaum, D.~Maynard,
  E.~Motta, and K.~Thirunarayan, editors, {\em The Semantic Web - ISWC 2009},
  volume 5823 of {\em Lecture Notes in Computer Science}, pages 293--309.
  Springer Berlin Heidelberg, 2009.

\bibitem{DBLP:conf/gvd/Koschmieder10}
A.~Koschmieder.
\newblock Cost-based optimization of regular path queries on large graphs.
\newblock In W.-T. Balke and C.~Lofi, editors, {\em Grundlagen von
  Datenbanken}, volume 581 of {\em CEUR Workshop Proceedings}. CEUR-WS.org,
  2010.

\bibitem{DBLP:conf/ssdbm/KoschmiederL12}
A.~Koschmieder and U.~Leser.
\newblock Regular path queries on large graphs.
\newblock In A.~Ailamaki and S.~Bowers, editors, {\em SSDBM}, volume 7338 of
  {\em Lecture Notes in Computer Science}, pages 177--194. Springer, 2012.

\bibitem{Ladwig:2010:LDQ:1940281.1940311}
G.~Ladwig and T.~Tran.
\newblock Linked data query processing strategies.
\newblock In {\em Proceedings of the 9th International Semantic Web Conference
  on The Semantic Web - Volume Part I}, ISWC'10, pages 453--469, Berlin,
  Heidelberg, 2010. Springer-Verlag.

\bibitem{Lim:2002:XOS:1287369.1287408}
L.~Lim, M.~Wang, S.~Padmanabhan, J.~S. Vitter, and R.~Parr.
\newblock Xpathlearner: An on-line self-tuning markov histogram for xml path
  selectivity estimation.
\newblock In {\em Proceedings of the 28th International Conference on Very
  Large Data Bases}, VLDB '02, pages 442--453. VLDB Endowment, 2002.

\bibitem{lynchDistributedAlgorithms}
N.~A. Lynch.
\newblock {\em {Distributed Algorithms}}.
\newblock Morgan Kaufmann, 1st edition, March 1997.

\bibitem{DBLP:journals/jodl/MendelzonMM97}
A.~O. Mendelzon, G.~A. Mihaila, and T.~Milo.
\newblock Querying the world wide web.
\newblock {\em Int. J. on Digital Libraries}, 1(1):54--67, 1997.

\bibitem{DBLP:conf/vldb/MendelzonW89}
A.~O. Mendelzon and P.~T. Wood.
\newblock Finding regular simple paths in graph databases.
\newblock In P.~M.~G. Apers and G.~Wiederhold, editors, {\em VLDB}, pages
  185--193. Morgan Kaufmann, 1989.

\bibitem{Nav04jstat}
G.~Navarro.
\newblock Pattern matching.
\newblock {\em Journal of Applied Statistics}, 31(8):925--949, 2004.
\newblock Special issue on Pattern Discovery.

\bibitem{Nejdl02edutella}
W.~Nejdl, B.~Wolf, C.~Qu, S.~Decker, M.~Sintek, A.~Naeve, M.~Nilsson,
  M.~Palm\'{e}r, and T.~Risch.
\newblock {EDUTELLA}: a {P2P} networking infrastructure based on {RDF}.
\newblock In {\em WWW '02: Proceedings of the 11th international conference on
  World Wide Web}, pages 604--615, New York, NY, USA, 2002. ACM.

\bibitem{plake2006alibaba}
C.~Plake, T.~Schiemann, M.~Pankalla, J.~Hakenberg, and U.~Leser.
\newblock Alibaba: Pubmed as a graph.
\newblock {\em Bioinformatics}, 22(19):2444--2445, 2006.

\bibitem{DBLP:journals/corr/RakhmawatiUKHH13}
N.~A. Rakhmawati, J.~Umbrich, M.~Karnstedt, A.~Hasnain, and M.~Hausenblas.
\newblock Querying over federated {SPARQL} endpoints - {A} state of the art
  survey.
\newblock {\em CoRR}, abs/1306.1723, 2013.

\bibitem{DBLP:journals/tcs/ShoaranT09}
M.~Shoaran and A.~Thomo.
\newblock Fault-tolerant computation of distributed regular path queries.
\newblock {\em Theor. Comput. Sci.}, 410(1):62--77, 2009.

\bibitem{DBLP:conf/vldb/Suciu96}
D.~Suciu.
\newblock Query decomposition and view maintenance for query languages for
  unstructured data.
\newblock In T.~M. Vijayaraman, A.~P. Buchmann, C.~Mohan, and N.~L. Sarda,
  editors, {\em VLDB}, pages 227--238. Morgan Kaufmann, 1996.

\bibitem{umbrichLinkedDataQuery}
J.~Umbrich, A.~Hogan, A.~Polleres, and S.~Decker.
\newblock Link traversal querying for a diverse web of data.
\newblock {\em Semantic Web Journal}, 2014.

\bibitem{Vardi:1982:CRQ:800070.802186}
M.~Y. Vardi.
\newblock The complexity of relational query languages (extended abstract).
\newblock In {\em Proceedings of the Fourteenth Annual ACM Symposium on Theory
  of Computing}, STOC '82, pages 137--146, New York, NY, USA, 1982. ACM.

\bibitem{verborgh2014low}
R.~Verborgh, O.~Hartig, B.~De~Meester, G.~Haesendonck, L.~De~Vocht,
  M.~Vander~Sande, R.~Cyganiak, P.~Colpaert, E.~Mannens, and R.~Van~de Walle.
\newblock Low-cost queryable linked data through triple pattern fragments.
\newblock In {\em Proceedings of the 13th International Semantic Web
  Conference: Posters and Demos}, 2014.

\bibitem{DBLP:journals/corr/YakovetsGG15}
N.~Yakovets, P.~Godfrey, and J.~Gryz.
\newblock Towards query optimization for {SPARQL} property paths.
\newblock {\em CoRR}, abs/1504.08262, 2015.

\end{thebibliography}

\end{document}